\documentclass[aps,prl, showpacs, reprint,superscriptaddress]{revtex4-1}

\usepackage{graphicx}
\usepackage{amssymb,amsmath}
\usepackage{bm}
\usepackage{float}
\usepackage{xcolor}
\definecolor{midnightblue}{cmyk}{1,1,0,0.1}
\definecolor{forestgreen}{cmyk}{0.75,0,1,0.5}

\usepackage{hyperref}
\hypersetup{
    bookmarks=true,         
    unicode=false,          
    pdftoolbar=true,        
    pdfmenubar=true,        
    pdffitwindow=false,     
    pdfstartview={FitH},    
    pdftitle={My title},    
    pdfauthor={Author},     
    pdfsubject={Subject},   
    pdfcreator={Creator},   
    pdfproducer={Producer}, 
    pdfkeywords={keyword1} {key2} {key3}, 
    pdfnewwindow=true,      
    colorlinks=true,       
    linkcolor=midnightblue,          
    citecolor=magenta,        
    filecolor=midnightblue,      
    urlcolor=midnightblue,          
}

\begin{document}

\title{The edge engineering of topological Bi(111) bilayer}

\author{Xiao Li}
\altaffiliation{Equal contribution.}
\affiliation{International Center for Quantum Materials, School of Physics, Peking University, Beijing 100871, China}
\affiliation{Collaborative Center for Quantum Materials, Peking University, Beijing, China}

\author{Hai-Wen Liu}
\altaffiliation{Equal contribution.}
\affiliation{International Center for Quantum Materials, School of Physics, Peking University, Beijing 100871, China}
\affiliation{Collaborative Center for Quantum Materials, Peking University, Beijing, China}

\author{Hua Jiang}
\affiliation{Department of Physics and Jiangsu Key Laboratory of Thin Films, Soochow University, Suzhou 215006, China}

\author{Fa Wang}
\affiliation{International Center for Quantum Materials, School of Physics, Peking University, Beijing 100871, China}
\affiliation{Collaborative Center for Quantum Materials, Peking University, Beijing, China}

\author{Ji Feng}
\email{jfeng11@pku.edu.cn}
\affiliation{International Center for Quantum Materials, School of Physics, Peking University, Beijing 100871, China}
\affiliation{Collaborative Center for Quantum Materials, Peking University, Beijing, China}

\date{\today}

\begin{abstract}
A topological insulator is a novel quantum state, characterized by symmetry-protected non-trivial edge/surface states. Our first-principle simulations show the significant effects of the chemical decoration on edge states of topological Bi(111) bilayer nanoribbon,
which remove the trivial edge state and recover the Dirac linear dispersion of topological edge state.  
By comparing the edge states with and without chemical decoration, the Bi(111) bilayer nanoribbon offers a simple system for assessing conductance fluctuation of edge states. The chemical decoration can also modify the penetration depth and the spin texture of edge states.
A low-energy effective model is proposed to explain the distinctive spin texture of Bi(111) bilayer nanoribbon, 
which breaks the spin-momentum orthogonality along the armchair edge.

\end{abstract}


\maketitle

As an insulating state with symmetry-protected gapless interface
 electronic modes, the topological insulator (TI) has received
considerable attention recently \citep{Hasan10,Qi11,Moore10}.
The edge conduction channels of two-dimensional (2d) TI exhibits quantum spin Hall effect within bulk gap \citep{Kane05a}.
A single bilayer Bi(111) film has been predicted to be a 2d TI with a large
 band gap of about 0.5 eV \citep{Murakami06, Wada11, Liu11a}, while other 2d TIs, 
 such as HgTe/CdTe quantum wells \citep{Konig07} and
InAs/GaSb quantum wells \citep{Knez12}, have  gaps of only several
tens of meV at best. Recently, Bi(111) bilayer has been readily grown on Bi$_{2}$Te$_{3}$ or Bi$_{2}$Se$_{3}$
substrates \citep{Hirahara11,Hirahara12,Yang12,Fukui12,Miao13,Wang13}. 
Therefore, it is very promising for room-temperature TI-based devices.
However, the native edges of Bi bilayer suffer from the
simultaneous presence of both trivial and non-trivial edge modes \citep{Wada11,Yang12},
which complicate the fundamental investigation of its topological properties and eventual
applications. Similar complication had perplexed the interpretation of surface states
in three-dimensional (3d) TI Bi$_{1-x}$Sb$_{x}$ \citep{Fu07,Hsieh08}.
Although localization in Anderson's sense will suppress
trivial conducting channels, quantitively localizing trivial channels
will still be an experimental challenge. A wide distribution of conductance
induced by multiple edge states is not desirable for accurate transport
measurement \citep{Li09a,Wada11,Sabater13}.

The complicated edge or surface states may be a generic problem associated with dangling bond
states at the termination of 2d or 3d TIs. In this Letter,
we report a first-principle analysis of chemical decoration
of the edge states of Bi(111) bilayer, which, as we show, is an effective
route for precise engineering of conducting edge states. We demonstrate that
chemical passivation quantitatively removes the trivial edge bands
in Bi bilayer nanoribbons, restoring the desired Dirac dispersion
of the non-trivial edges.  We further compute transport and optical signatures of the chemical
decoration of the Bi bilayer edges, which can be assessed experimentally.
In particular, we suggest that the Bi bilayer nanoribbons, with and
without chemical passivation, offer a simple system for assessing
conductance fluctuation of edge states. Moreover, edge decoration has important
consequences on the spatial distribution and spin texture of the edge
states.  A low-energy effective model is proposed 
to explain the distinctive spin texture of Bi bilayer nanoribbon, 
 where the spin is no longer perpendicular to the momentum along the armchair edge.

\begin{figure}[b!]
\includegraphics[width=7.5 cm]{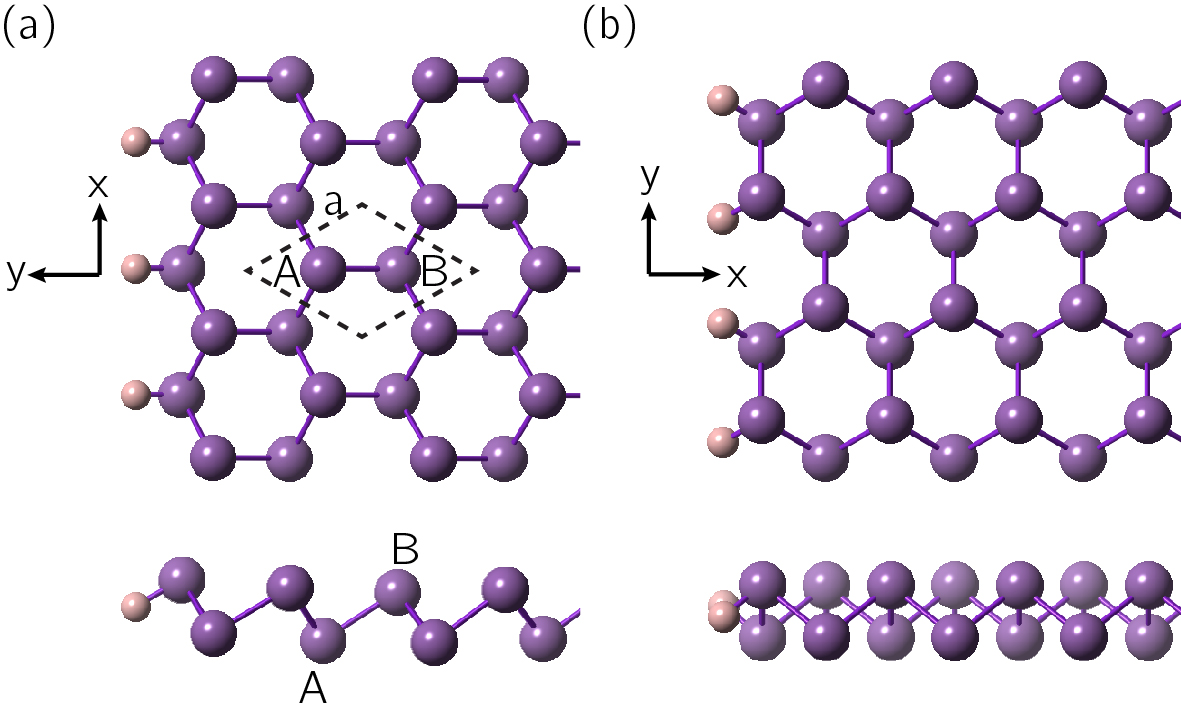}
\caption{The geometric structure of (a) the zigzag and (b) the armchair Bi(111) nanoribbon.
Up: The top view. Down: The side view. 
The primitive cell of single bilayer sheet, 
bounded by dashed lines, are also shown. 
The zigzag and armchair edges are perpendicular to each
other, which are along x- and y-axis of the Bi(111) sheet, respectively.
The hydrogen-terminated edges shown in the figure are the most stable
structure of hydrogen adsorption. The big purple ball stands for bismuth
atom, and the small pink ball for hydrogen atom. Idoine-terminated
nanoribbons have the similar structures and they are not shown.}
\label{fig:structure}
\end{figure}

We use density functional theory \citep{Kohn65} calculations implemented
in the Vienna Ab initio simulation package \citep{Kresse96,Kresse99}
to investigate geometric and electronic structure of single Bi(111) bilayer
and its nanoribbons. Computational details can be found in the Supplementary Information (S.I.) \citep{SI}. 
Notice that spin-orbit coupling (SOC) is included in the calculation of electronic structure, unless otherwise specified.
The single Bi(111) bilayer has the point group symmetry of $D_{3d}$ with spatial inversion included.
As shown in Fig.~\ref{fig:structure}, the top view of Bi(111) bilayer shows a bipartite honeycomb lattice with A and B sublattices. 
Two sublattices have different heights, forming bilayer structure.
The calculated nearest-neighbor bond angles in single Bi bilayer is
91$^\circ$ and the lattice constant $a=4.34$ \AA.
Based on the electronic structure of Bi bilayer (See S.I. \citep{SI}), the band inversion 
takes place between $p$-like valence and conduction bands at $\Gamma$ point, 
 leading to a 2d topological insulator with a indirect band gap of 0.5 eV, agreeing with the previous results \citep{Murakami06, Wada11,Liu11a}.

\begin{figure}[t]
\includegraphics[width=7.5 cm]{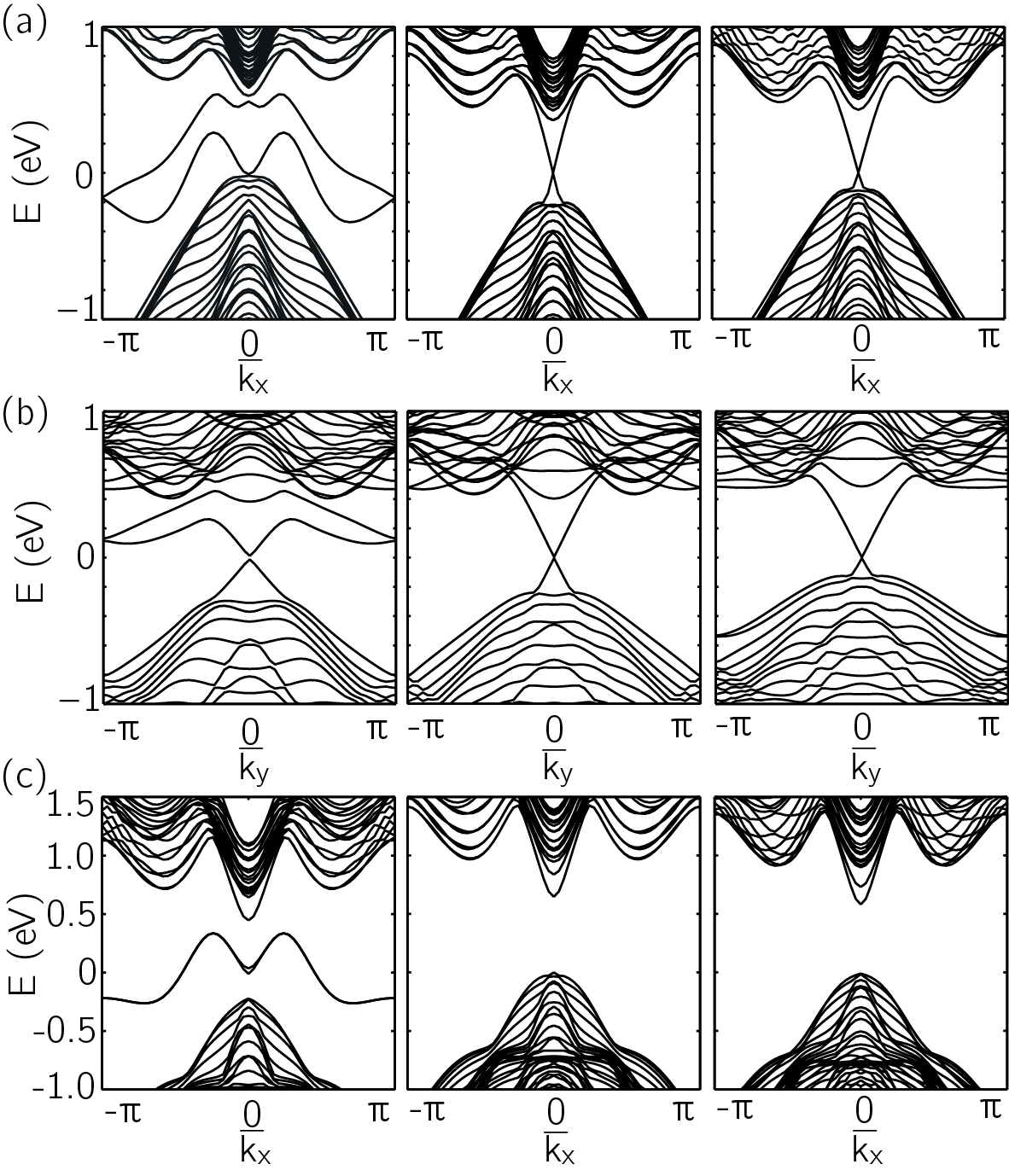}
\caption{The band structure of Bi(111) nanoribbons. 
(a) The zigzag nanoribbons with SOC. 
(b) The armchair nanoribbons with SOC.
(c) The zigzag nanoribbons without SOC. The armchair nanoribbons without
SOC are shown in Fig. S3. There are three panels in each figure.
Left panel: The native edge. Middle panel:
The hydrogen-terminated edge. Right panel: The iodine-terminated
edge. 
The highest occupied energy level is set to zero energy.
$\overline{k}_{x}\equiv k_{x}a$
and $\overline{k}_{y}\equiv\sqrt{3}k_{y}a$.}
\label{fig:band}
\end{figure}

To investigate the edge properties of the Bi(111) nanoribbons, we
study two representative model systems: (1) a 40-atom (per unit cell)
zigzag nanoribbon (about 7.3 nm wide), and (2) 50-atom (per unit cell)
armchair nanoribbon (about 5.2 nm wide) (Fig.~\ref{fig:structure}).
For nanoribbons with native edges,
 the band structures are shown in the left panels of Figs.~\ref{fig:band} (a) and (b). 
 All bands remain spin-degenerate, 
 owing to simultaneous time-reversal and inversion symmetry.
  Within the bulk band gap, the upper non-trivial and the lower trivial edge
states are present simultaneously, which span the entire Brillouin zone (BZ).
There are odd numbers of Kramers pairs of edge states
at the Fermi level, showing that Bi(111) bilayer is indeed a 2d topological
insulator. If we sweep the chemical potential across the gap by external
gating, the number of conducting channel may change from 3 to 1 (or
from 1 to 3). Here, the edge states do not show linear dispersion near $\Gamma$ point,
 in contrast to Kane-Mele model of quantum spin Hall effect\citep{Kane05a}.

Given that the atomic edge adsorption of graphene nanoribbon has been
achieved via hydrogen plasma etching recently and 
the edge decoration has important effects on electronic properties of
graphene nanoribbons \citep{Kan08,Loh10,Zhang12a}, 
the edge states of Bi(111) nanoribbon may be modified by chemical adsorption. 
We study the adsorption of hydrogen and
iodine atoms on Bi(111) zigzag and armchair nanoribbons. Considering
different adsorption sites from the edge to the middle part of nanoribbons (Fig. S2 in S.I. \citep{SI}),
the atom adsorption on the outmost bismuth atom is the most stable
structure with the lowest adsorption energy \citep{SI}, compared with atom adsorption on the basal plane. It will lead to selective edge decoration in experiment, similar
with graphene nanoribbon \citep{Zhang12a}. The adatoms restore three-fold coordination of the bismuths at two edges, indicating that the dangling bonds become saturated.

The middle and right panels in Figs.~\ref{fig:band} (a) and (b) show the band structure after the edge functionalization for zigzag and armchair
nanoribbons, respectively. Compared with native nanoribbons,
there are only linear dispersing non-trivial edge states in the center
of BZ, while the trivial edge states are removed. The Fermi
velocities are $8.5\times10^{5} m/s$ and $7.9\times10^{5} m/s$ for
hydrogen- and iodine-terminated zigzag nanoribbon, respectively.
And the values are $7.7\times10^{5} m/s$ and $7.3\times10^{5} m/s$
for hydrogen- and iodine-terminated armchair nanoribbon, respectively. 

The band structures Fig.~\ref{fig:band}(c) and Fig. S3 further show the corresponding band structure of Bi(111) nanoribbon without considering SOC.
For the native edge, the edge states within the bulk gap are trivial ones in the absence of SOC.
With atom adsorption at two edges, the trivial edge states are removed from the bulk gap. Therefore, taking into account SOC in our calculation, 
the emergent edge states (the middle and right panels in Figs.~\ref{fig:band}(a) and (b)) after chemical decoration are only non-trivial ones, resulting from the band inversion of TI.

\begin{figure}[t]
\includegraphics[width=6.5 cm]{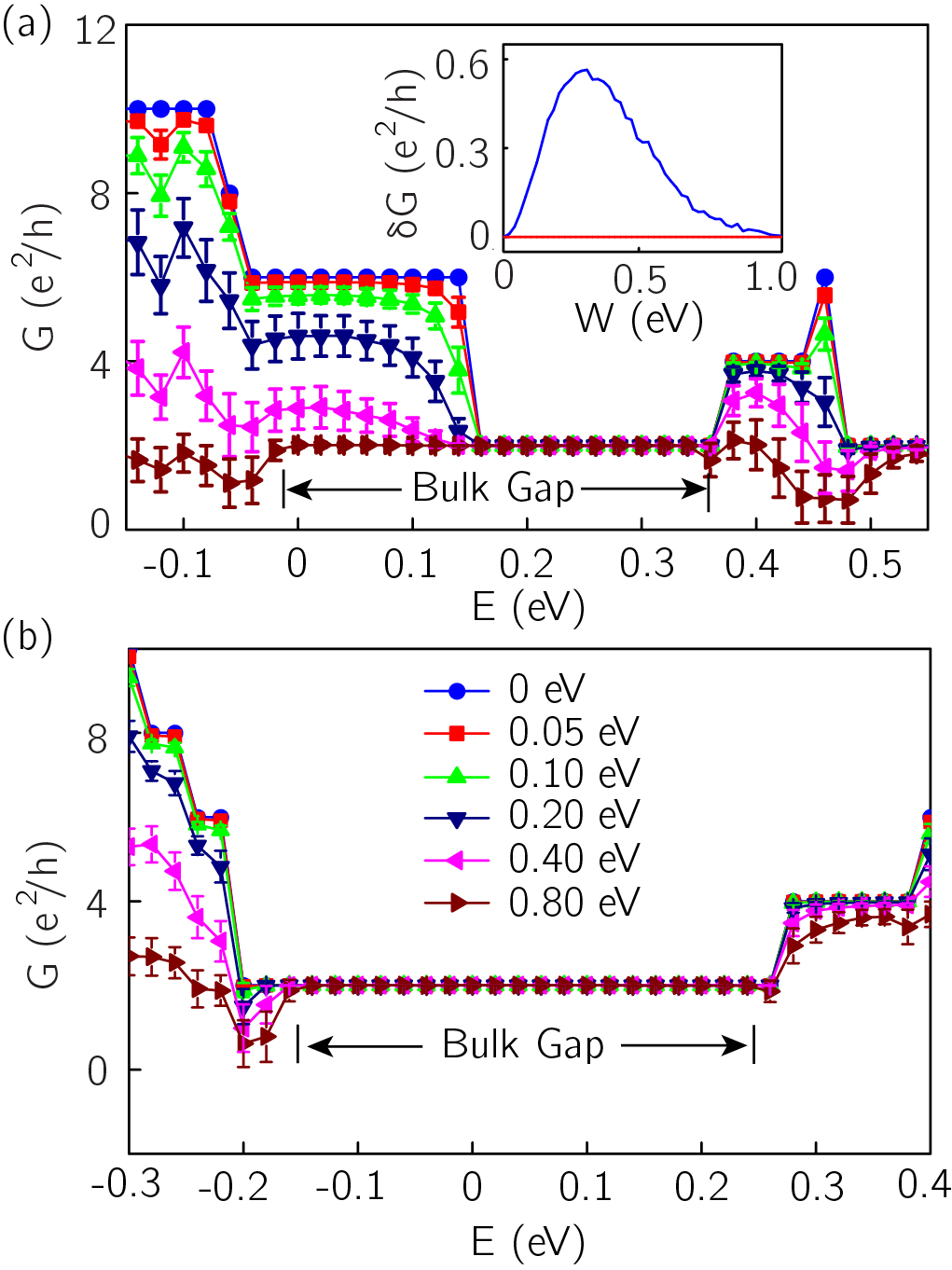}
\caption{Conductance, $G$, as the shift of the chemical potential for (a) the native  and 
(b) hydrogen-terminated zigzag nanoribbon.
Under Anderson disorders of different
strengths $W$, the conductances of the nanoribbons are shown in different colors and symbols. The bar
on every data point represents the conductance fluctuation, $\delta{G}$. The inset
of (a) shows the conductance fluctuation as a function of $W$. The Fermi level is set to zero energy. The blue line stands for the edge states at 0.05
eV. The red line stands for the ones at 0.25 eV, which keeps zero.}
\label{fig:tran}
\end{figure}

Considering the significant modification of  the edge bands by chemisorption, we suggest the effects of the edge engineering 
can be directly probed by transport and optical measurements. 
The key effect of the edge chemisorption is turning the number of edge conduction channels
from three in the native nanoribbon to only one. This creates an interesting
experimental apparatus to assess the effects of localization in the
Anderson's paradigm. We expect that in the case of single non-trivial
edge channel, the conduction will stay quantized and will not be affected
by non-magnetic disordered Anderson scatterers. On the other hand,
the simultaneous presence of both non-trivial and trivial edge channels
will show rather different transport behavior. Sufficiently strong
disorder will eventually localize the trivial channels. However, in the
intermediate localization regime, we may have window to detect disorder-induced
conductance fluctuation \citep{Li09a}. The edge modification will also change the optical absorption of the material,
which can also be measured experimentally. We therefore compute
the transport and optical spectra of Bi(111) bilayer nanoribbons, based
on a full-valence tight-binding (TB) model from DFT calculations \citep{SI}.

Taking Bi(111) zigzag nanoribbon as an example, the transport spectra are calculated with the non-equilibrium Green's
function approach \citep{Meir92}. For the native zigzag nanoribbon, 
the conductance is $6e^{2}/h$ at the Fermi level,
as expected and also consistent with experimental measuration \citep{Sabater13}.
The conductance of edge states changes from $6e^{2}/h$ to $2e^{2}/h$
as the chemical potential is gated up within the bulk band gap, leaving
only the contribution from the non-trivial edge state. Upon introduction
of Anderson disorder to the model, the conductance from trivial edge
gradually decays with the increasing strength of disorder in the neighboring
of the Fermi level (Fig.~\ref{fig:tran}(a)), indicating localization.
At the same time, we indeed observe significant conductance fluctuation, $\delta{G}$. 
We see that $\delta{G}$ first increases with weak Anderson disorder, 
but eventually decreases to zero upon complete localization of trivial edge channels 
(the inset of Fig.~\ref{fig:tran}(a)). For hydrogen-termination zigzag nanoribbon,
the conductance stays at the quantized platform of $2e^{2}/h$ without
any fluctuations within bulk band gap, as shown in Fig.~\ref{fig:tran}(b). We also investigate
the effect of less-than-full adsorption along two edges of nanoribbon
with random occupancy of the adsorbed hydrogen atoms. It is found
that 25\% coverage of the edges with hydrogen adsorption can obtain
a similar conductance plateau of $2e^{2}/h$ as the one of 100\% hydrogen
adsorption, showing that it is feasible to improve the transport properties
by partial edge decoration.

\begin{figure}[t]
\includegraphics[width=7 cm]{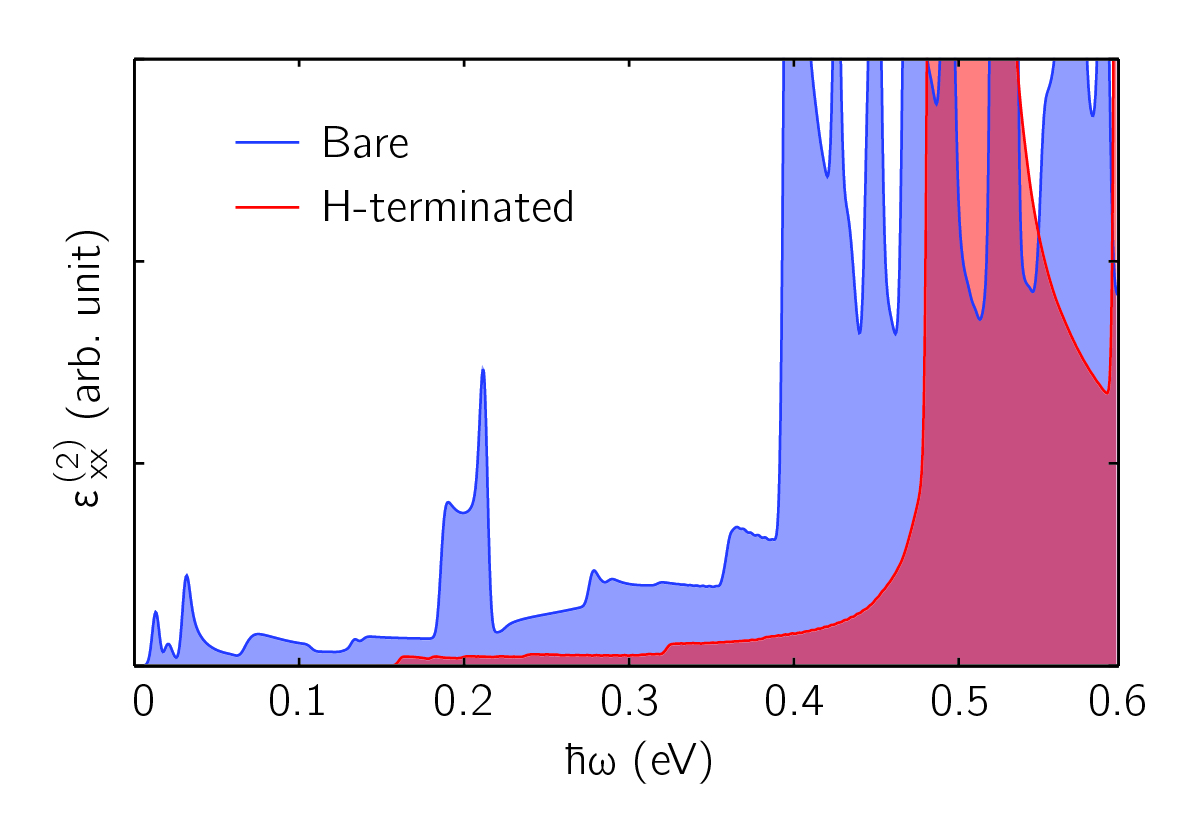}
\caption{The imaginary part of the dielectric tensor component,
$\varepsilon_{xx}^{(2)}(\omega)$, as a function of the optical frequency $\omega$
for zigzag nanoribbon. The bare edge and the hydrogen-terminated edge
are shown in blue and red , respectively.}
\label{fig:optical}
\end{figure}

Fig.~\ref{fig:optical} shows the imaginary part of the dielectric tensor component for zigzag
nanoribbon, $\varepsilon_{xx}^{(2)}(\omega)$, under the $x$-polarized irradiation field. Without any terminations, nonzero dielectric function
shows that there are alway optical transitions between edge states
or between edge states and bulk states. However, these
transitions are inhibited to a great extent for hydrogen-terminated
zigzag nanoribbon. These observable characteristics can be used as
signals of the edge decoration of Bi nanoribbons. 

\begin{figure}[t]
\includegraphics[width=8 cm]{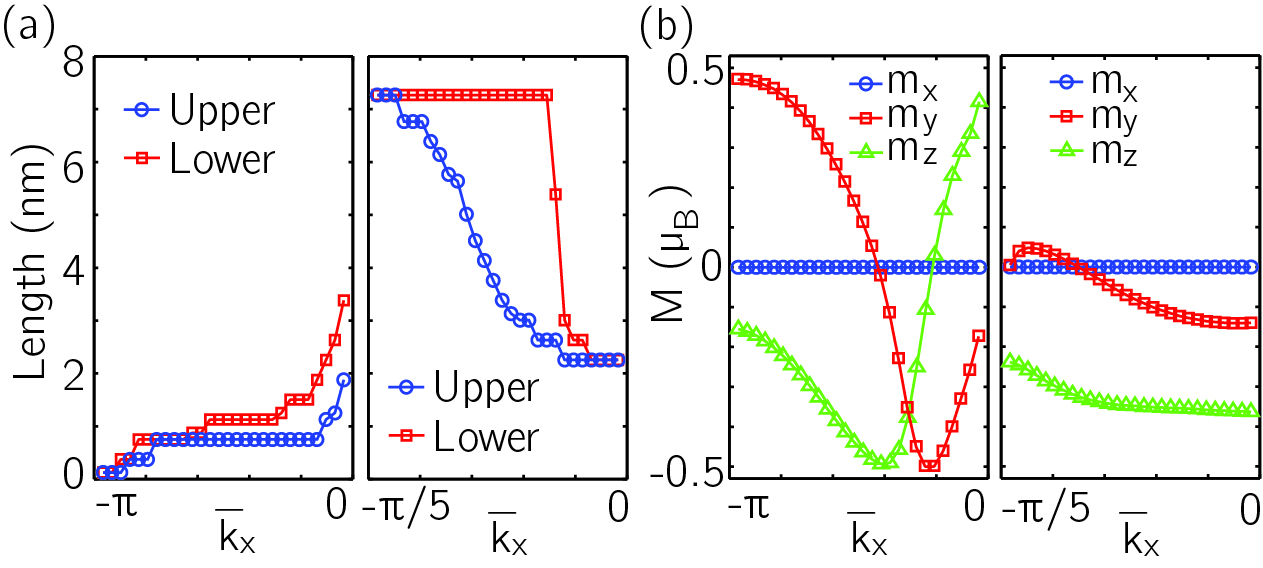}
\caption{The electronic properties of Bi(111) zigzag edge states. 
(a) The penetration depth of the edge states as a function of the momentum $k_{x}$. Left: The
native edge. Right: The hydrogen-terminated edge. The upper and lower
edge states are shown by blue circle and red square, respectively.
(b) The spin texture of edge states. The blue circle, red square and green triangle
stand for three components of spin moment, $m_{x}$, $m_{y}$ and
$m_{z}$, respectively. $\overline{k}_{x}\equiv k_{x}a$. In (b), Only the more localized edge state (the upper edge) are shown,
while the other one behaves in the same manner.}
\label{fig:spin}
\end{figure}

Alongside with the change of the band structure and the corresponding experimental signatures, 
the penetration depth and spin texture of edge states have been modified by edge engineering.
Fig.~\ref{fig:spin}(a) shows the change of the penetration depth with
atomic adsorption for zigzag nanoribbon. For the zigzag nanoribbon with native edge,
only one band of each group of the spin-degenerate edge bands
are presented in the half of the BZ ($k_{x}a$ from $-\pi$ to 0), while
other states can be obtained by the inversion and time-reversal symmetry.
 For hydrogen-terminated nanoribbon, the edge states are presented in the smaller zone of the BZ
($-\pi/5$ to 0) to zoom in on the linear dispersion. The penetration
depth is very short for both the upper non-trivial (blue circle in the left panel of
Fig. ~\ref{fig:spin}(a)) and lower trivial edge states (red square) of the 7.3 nm-width
nanoribbon without any termination.  For states away from the bulk states ($k_{x}a$ near $\pi$), 
they are localized within 1nm closest to the edge.
When the edge state approaches the bulk states ($k_{x}a$ near $0$), the penetration depth is gradually getting
longer. In contrast, the penetration depths of the hydrogen-terminated edge states are all 
more than 2 nm, much longer than the one of native edge, which agrees with the inverse relationship 
between the penetration depth and the momentum-space
width of the edge state \citep{Volkov85,Wada11}. Besides, the penetration depths is also getting longer for hydrogen-terminated edge,
as the edge state approaches the bulk states ($k_{x}a$ from $0$ to $\pi/5$).

The spin moments, $m_{x}$, $m_{y}$
and $m_{z}$, for Bi(111) zigzag nanoribbon are shown in Fig.~\ref{fig:spin}(b). Compared with Kane-Mele model \citep{Kane05a}
and HgTe/CdTe quantum wells \citep{Konig07}, the edge states
of Bi(111) nanoribbons have more complicated spin textures. For native
zigzag nanoribbon, the components, $m_{y}$ and $m_{z}$,
perpendicular to the momentum $k_{x}$ of zigzag edge, vary
gradually and the spin direction rotates with the momentum. The component
$m_{x}$, parallel to the momentum, is zero. It is similar to the interface state of well-known topological insulators, such as the natural cleavage (111) surface of Bi$_{2}$Se$_{3}$, where the spins are locked to their momentums at right angles \citep{Hasan10}.
For hydrogen-terminated zigzag edge, $m_{y}$ and $m_{z}$ have altered their trends and varied
slowly in the momentum space near the Dirac point, with
$m_{x}=0$. The armchair edges have  similar spin textures with zigzag edges, 
where $m_{x}$ is still zero, as shown in Fig. S4 \citep{SI}.
 Given that the momentum direction is  along y axis for armchair edges,
 the spin is no longer perpendicular to the momentum, 
 in contrast with the spin-momentum orthogonality in well-known TIs \citep{Hasan10,Kane05a,Konig07}. 
 This departure can be explained by our effective model of single Bi(111) bilayer.

We suggest a low-energy effective Hamiltonian of single Bi(111)
bilayer based on the symmetry analysis. Considering the inversion
symmetry of single bilayer, we combine the $p$-orbitals near the
Fermi level to form the bonding and anti-bonding states with definite
parity, $\left|p_{\lambda}^{\pm}\right\rangle =\frac{1}{\sqrt{2}}\left(\left|p_{A,\lambda}\right\rangle \mp\left|p_{B,\lambda}\right\rangle \right)$,
where $p_{\lambda}=p_{x,y,z}$ stand for three $p$-orbitals and
A/B for A/B sublattice stands of the honeycomb lattice. The superscript
$\pm$ correspond to even and odd parity, respectively. Taking into
account the band splitting from both crystal field and SOC, the band inversion
mainly arises between degenerate states $\left|p_{z}^{-},\pm\frac{1}{2}\right\rangle $
and degenerate states $\left|p_{x,y}^{+},\pm\frac{3}{2}\right\rangle $ near $\Gamma$,
where $\pm\frac{1}{2}$ and $\pm\frac{3}{2}$ denote the corresponding
total azimuthal quantum numbers.

We then construct four Wannier
bases, $|\alpha^{-}\rangle,$ $|\beta^{+}\rangle,$ $\hat{\mathcal{T}}|\alpha^{-}\rangle,$
and $\hat{\mathcal{T}}|\beta^{+}\rangle$, to describe the low-energy excitations of single
Bi(111) bilayer, where $|\alpha^{-}\rangle = |p_{z}^{-},\frac{1}{2}\rangle$ and $|\beta^{+}\rangle =N_{0}\left(\left|p_{x,y}^{+},\frac{3}{2}\right\rangle +\eta\left|p_{x,y}^{+},-\frac{3}{2}\right\rangle \right)$. $\hat{\mathcal{T}}$
is time-reversal operator, $N_{0}$ is the normalization factor and $\eta$ is the weight factor (See details in S.I. \citep{SI}).
The effective Hamiltonian near $\Gamma$
is separated into two subblocks as below,

\begin{equation}
\mathcal{H}({\bf k})=\left[\begin{array}{cc}
H({\bf k}) & 0\\
0 & H^{*}(-{\bf k})
\end{array}\right],
\end{equation}

{\small 
\begin{equation}
H(\mathbf{k})=m[\sigma_{z}+k^{2}(\lambda^{2}\sigma_{0}-\xi{}^{2}\sigma_{z})]+\hbar v(k_{x}\sigma_{x}-k_{y}\sigma_{y}),
\end{equation}
}where  $\sigma_{i}$ ($i=x,y,z,0$) is the Pauli matrices addressing the subspace
spanned by $|\alpha^{-}\rangle$ and $|\beta^{+}\rangle.$ The parameters
$m = 0.291$ eV, $\lambda = 14.11$ \AA, $\xi = 15.51$ \AA,
and $v = 1.079\times10^{6}\,\text{m/s}$,  which are obtained by fitting the DFT
band structure of Bi(111) bilayer near $\Gamma$
point, as shown in Fig. S1 \citep{SI}. The Hamiltonian leads to topological edge states with clean linear
dispersion and a Fermi velocity of $6.0\times10^{5}\,\text{m/s}$,
agreeing with our first-principle results. Moreover,
when spin Pauli matrix $s_{x}$ acts on the bases, we have $\left\langle \varphi\right|s_{x}\left|\varphi\right\rangle =0$,
where $\varphi=\alpha^{-},\beta^{+}$. That is, the low-energy bulk
bands have vanishing $m_{x}$, and so do the topological edge
states of both zigzag and armchair edges, which arise from the bulk band inversion. 
In contrast,  nonzero $m_{y}$ and $m_{z}$ can also be obtained by the Hamiltonian \citep{SI}.
This explains the distinctive spin-momentum relationship in Bi(111) bilayer.

In conclusion, the edge chemical decoration of Bi(111) bilayer
can significantly modify topological edge states. The experimental signatures and the low-energy effective
Hamiltonian are also proposed. Clean edge
state and model Hamiltonian will facilitate further investigations on
topological properties of the Bi(111) bilayer, such as superconducting proximity effect \citep{Fu08} and
topological Anderson Insulator \citep{Li09b}.

\textcolor{forestgreen}{\emph{\textsf{Acknowledgements}.}}---
This work is supported by the National Science Foundation of China under Grants Nos. 11174009 and 11374219, by China 973 Program Projects 2013CB921900.

\end{document}